\documentclass[a4paper,11pt]{article}

\usepackage{amsmath}
\usepackage{amsfonts}
\usepackage[latin1]{inputenc}
\usepackage{graphicx}
\usepackage[blocks]{authblk}
\usepackage[font=small,labelfont=bf]{caption}
\usepackage[colorlinks=true, urlcolor=blue, linkcolor=green, citecolor=red]{hyperref}
\usepackage[numbers,sort&compress]{natbib}
\usepackage{doi}

\providecommand{\keywords}[1]{\textbf{\small \textit{keywords:}} {\small #1}}

\title{\sc A Differential Phase Shift Scheme for Quantum Key Distribution in Passive Optical Networks}

\author{Michael Hentschel\thanks{\href{mailto:michael.hentschel@ait.ac.at}{michael.hentschel@ait.ac.at}}}
\author{Andreas Poppe}
\author{Bernhard Schrenk}
\author{Momtchil Peev}
\author{\\Edwin Querasser}
\author{Roland Lieger}
\affil{\small Digital Safety \& Security Department, AIT Austrian Institute of Technology GmbH, \\Donau-City-Strasse 1, 1220 Vienna, Austria}

\begin{document}

\maketitle

\begin{abstract}
\normalsize
\noindent
We propose a scheme for quantum key distribution (QKD) in a passive optical network (PON) based on differential phase shift (DPS) coding. A centralized station including all expensive components serves many users, making it suitable for a true multi-user network in a local environment with moderate distances on the order of a few kilometers. The emphasis lies on an asymmetric design for cost effective implementation of network end points.
\end{abstract}


\section{Introduction}
\label{sec:intro}
Quantum key distribution (QKD) has evolved from a pure research topic to a mature technology in recent years, playing an increasingly important role in secure communications. There are already a few commercial products on the market, however, broad acceptance in the community is still rather limited. One reason for this shall be described in the following.
\par
In classical cryptography currently used for electronic communications, purely algorithmic methods (SSH, TLS and the like) are employed to establish a secure connection between any two communication parties across an arbitrarily large network (i.e. the whole world wide web). It only requires the installation of the proper software on both sides without the need for special knowledge or dedicated hardware. With QKD the situation is entirely different. It deals with the transmission of photons on the lowest physical layer. Simply by its nature this technology strongly relies on (quantum) physical effects and is thus inherently hardware dependent. The hardware is usually quite sophisticated and in most cases rather expensive, at least far from being just a plug-in card for the computer. Moreover, most QKD systems are designed as point-to-point links and it is not straight forward how to combine them to larger and flexible networks. A few QKD networks have been demonstrated by connecting several individual systems in so-called trusted nodes \cite{Peev2009}, \cite{Sasaki2011}. But because none of the expensive hardware is shared among users, this approach does not improve the cost situation as compared to the sum of the individual links. Such networks may serve as a backbone infrastructure, but the service to multiple end users is not covered.
\par
In this paper we propose a scheme for an access multi-user QKD network with special attention on cost efficiency. The architecture is designed with strong asymmetry as to make network leafs as cheap as possible. This may bridge the ``last mile'' when connecting end users to a backbone network. We should mention that the system is intended for a local environment and moderate key rates. These might sound like severe limitations at first glance, but nonetheless we can think of several scenarios, where it may prove useful for securing small but crucial messages. Among these are office or government buildings, power plants, airport flight control and the like.
\par
In the following sections we will present our concept, the technical implementation and some test results of our demonstration setup.

\section{Concept and design}
\label{sec:concept}
\subsection{System architecture}
\label{subsec:arch}
As pointed out above, we strive for a system that is efficient in terms of cost per user. Certain expensive parts of some sort are unavoidable in all implementations, typically photon sources and single photon detectors. Our basic idea was to introduce a strong asymmetry, i.e. concentrate the inevitable expensive components in one single central unit and keep the multiple end units as cheap as possible. This concept gives rise to certain implications for the network topology.
\par
\begin{figure}[h]
\begin{center}
\includegraphics[height=50mm]{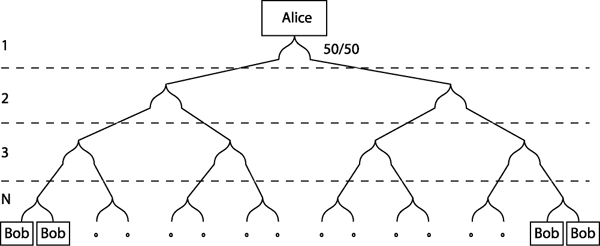}
\caption{A binary tree network topology with symmetric beam splitters. $N$ splitting levels provide $2^N$ leafs and require $2^N-1$ beam splitters.}
\label{fig:topos}
\end{center}
\end{figure}

From a topological point of view there are several possible network configurations such as star, tree or bus. Although not strictly necessary, it is preferable that the total transmission loss and the optical distance are similar for each network leaf, as we shall see later. Moreover, in order to keep the fiber network itself simple and the protocol overhead low, we chose to use a passive network made of beam splitters solely (as opposed to active switches). The above requirements are fulfilled best by a binary tree network with 50:50 beam splitters (Fig. \ref{fig:topos}).
\par
From our above demands arises another requirement concerning the system layout. Since they are expensive components, it is beneficial for us to have both the photon source and the detectors in the central unit. At first sight this sounds weird, as sender and receiver are usually at opposite sides. However, it is possible to break with this traditional setting, which leads us to what we call the ``central feeder'' concept. In short, the central unit (henceforth ``Alice'') produces classical photon pulses and distributes them to the network leafs (henceforth ``Bob'') via the fiber network. They, in turn, attenuate the pulses down to single photon level, modulate them according to a locally generated random sequence and send them back over the network. Finally Alice demodulates and detects the photons, thus obtaining (partially) the random sequences of the Bob units. A somewhat related concept was demonstrated in the so-called ``plug \& play'' system by idQuantique \cite{Ribordy1998}.

\subsection{Packet mode}
\label{subsec:packet}
The central feeder scheme described above implies the need for a pulse packet mode of operation. This becomes clear when considering the different power levels involved in the down-stream (towards Bob) and the up-stream (towards Alice) and the physical effects taking place on the fiber. Among these are the finite isolation of circulators, reflections from fiber connectors and Rayleigh backscattering \cite{Nakazawa1983}. Since the usable signals to be expected returning to Alice are on a sub-single photon level, they would plainly be overwhelmed by spurious photons. To amend this situation, we introduce pulse packets that travel back and forth through the network, returning to Alice in times of ``silence''. Additionally, the packets from different users can be interleaved, which makes a quasi-simultaneous operation possible without the need to address each Bob individually. The exact timing requirements and their connection with the network layout will be discussed in section \ref{subsec:downstream}.

\subsection{Coding scheme}
\label{subsec:coding}
\begin{figure}[h]
\begin{center}
\includegraphics[height=50mm]{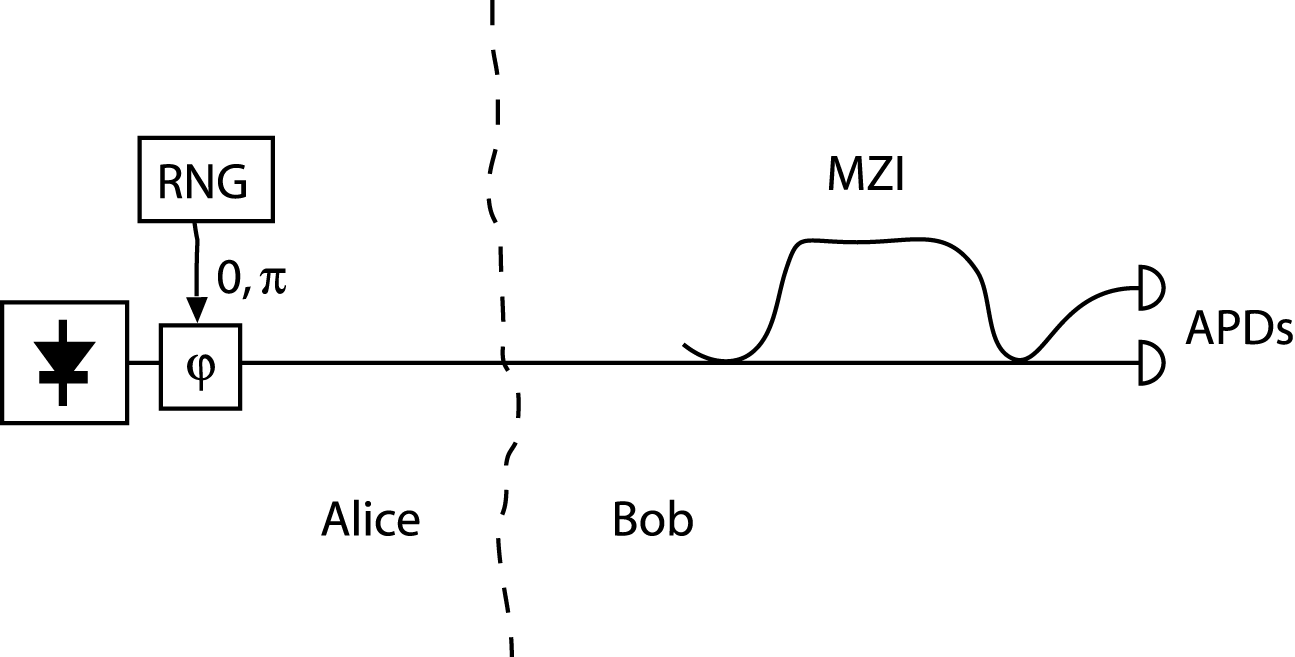}
\caption{Basic principle of differential phase shift coding, with phase modulator $\varphi$, random number generator RNG, Mach-Zehnder interferometer MZI and avalanche photo detectors APDs.}
\label{fig:dps}
\end{center}
\end{figure}

QKD systems have been demonstrated with a variety of coding schemes. The most prominent are polarization, phase and time-bin coding and continuous variables. All of them have their pros and cons, which we will not discuss in detail here. Anyway, we opted for an even different approach, namely differential phase shift (DPS) coding \cite{Diamanti2006a}. The basic principle is shown in figure \ref{fig:dps}.
The prerequisite for this technique is a pulsed laser source with a stable repetition rate and a coherence length spanning over many pulses. To every pulse a random but constant phase modulation is applied, selected from two possible values ($0$ or $\pi$) by a random number generator. On the receiver side an unbalanced Mach-Zehnder interferometer (MZI) performs the demodulation. To this end, each arriving pulse is split in two equal halves. The arms of the interferometer differ in propagation delay by exactly the repetition period of the laser pulses. In this manner, the halves of two adjacent pulses are superimposed on the second beam splitter. Since they initially had a fixed phase relation, they will emerge in one or the other output port, depending on their phase difference ($\Delta \varphi = 0$ or $\pm \pi$) applied by the phase modulator, respectively. Surprisingly enough, this scheme works not only for classical pulses but just the same for single photons and even for pulses with an average photon number of much less than unity.

\section{Technical implementation}
\label{sec:implement}
\subsection{Overview}
\label{subsec:overview}
\begin{figure}[h]
\begin{center}
\includegraphics[height=70mm]{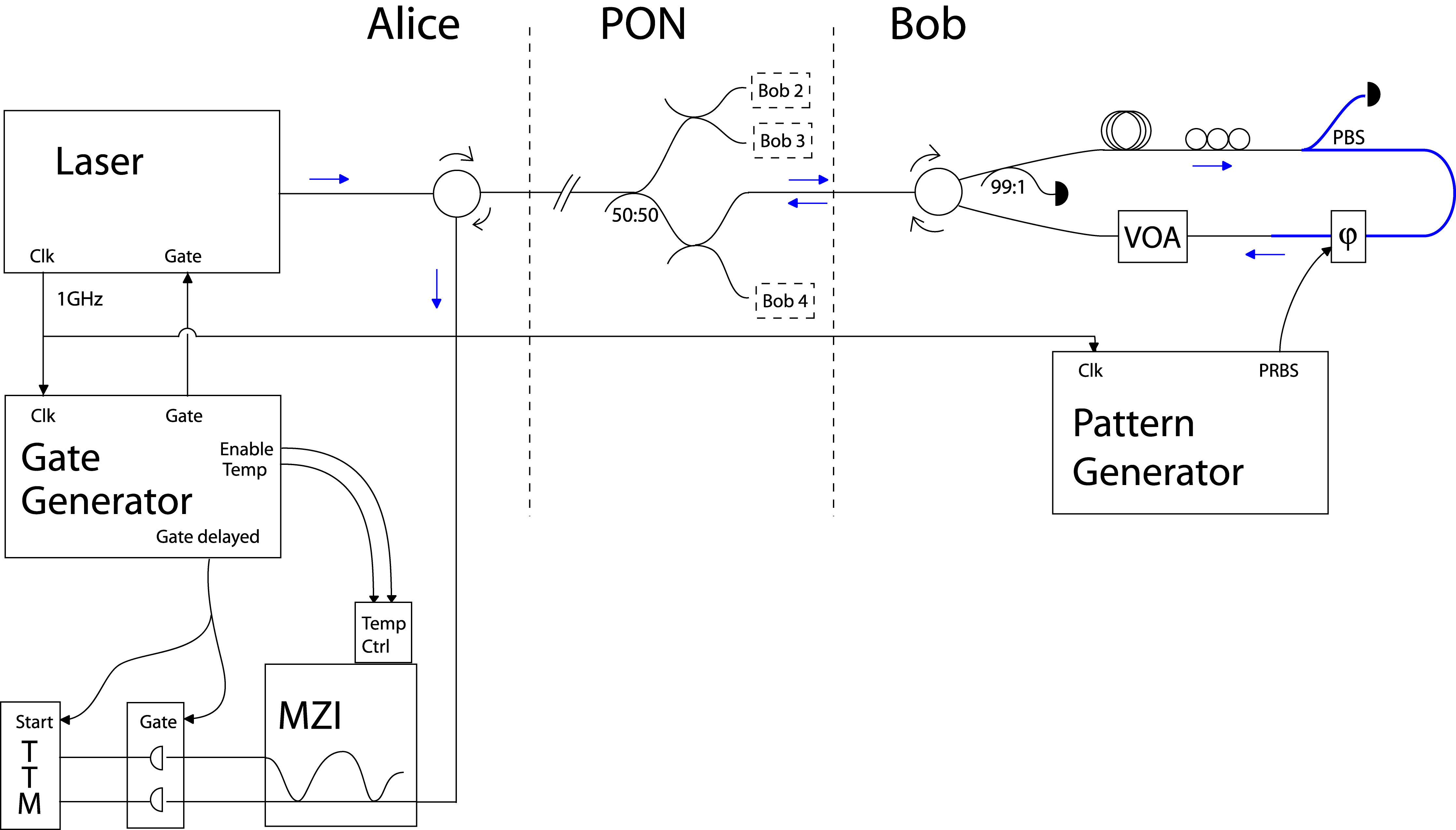}
\caption{Schematic of the setup.}
\label{fig:setup}
\end{center}
\end{figure}

Our complete setup is shown in figure \ref{fig:setup}. As an overall description, classical laser pulses are sent out from Alice and are distributed to multiple Bob units over the network. Bob modulates and attenuates the pulses and sends them back to Alice, where they are demodulated and detected. The individual sections shall be discussed now in detail.

\subsection{Photon source (Alice)}
\label{subsec:source}
As a primary laser source we use a telecom DFB laser module (\textit{3CN00410DT, 3S Photonics}). It emits a cw-power of $P_{cw} = 20$ mW at $\lambda = 1550.12$ nm with a spectral width of $\Delta \nu = 2$ MHz. This corresponds to a coherence length of $\tau_c = 500$ ns. The light is delivered through a PM pigtail which is connected to an intensity modulation system (\textit{ModBox, Photline Technologies}). The ModBox includes a master clock ($f_p = 1$ GHz), a pulse generator ($\tau_p = 120$ ps), a 10 GHz LiNbO$_3$ intensity modulator and an automatic bias control board. A gating input provides the means to generate packets of pulses. Although the modulator exhibits a relatively high static extinction ratio of 30 dB, we need to further suppress photons outside our pulse packets with a semiconductor optical amplifier (SOA) (\textit{OPB-12-15-N-C-FA, Kamelian}) that is driven by a copy of the gating signal. The so achieved background level is crucial, since it will be present as accumulated noise of all reflections from the network (see section \ref{subsec:downstream}).
\par
The Alice unit includes an electronic board, developed by AIT, that takes care of generating the pulse packets and all further synchronization thereto, plus additional controls like the MZI temperature. It is driven by the primary clock and provides a synchronized gate signal for the laser modulator along with some derived signals thereof for triggering the detectors and time tagging (which will be described later). The gate parameters are widely programmable including the packet length, the packet repetition period and the phase of the packet start.

\subsection{Down-stream (PON)}
\label{subsec:downstream}
\begin{figure}[h]
\begin{center}
\includegraphics[height=70mm]{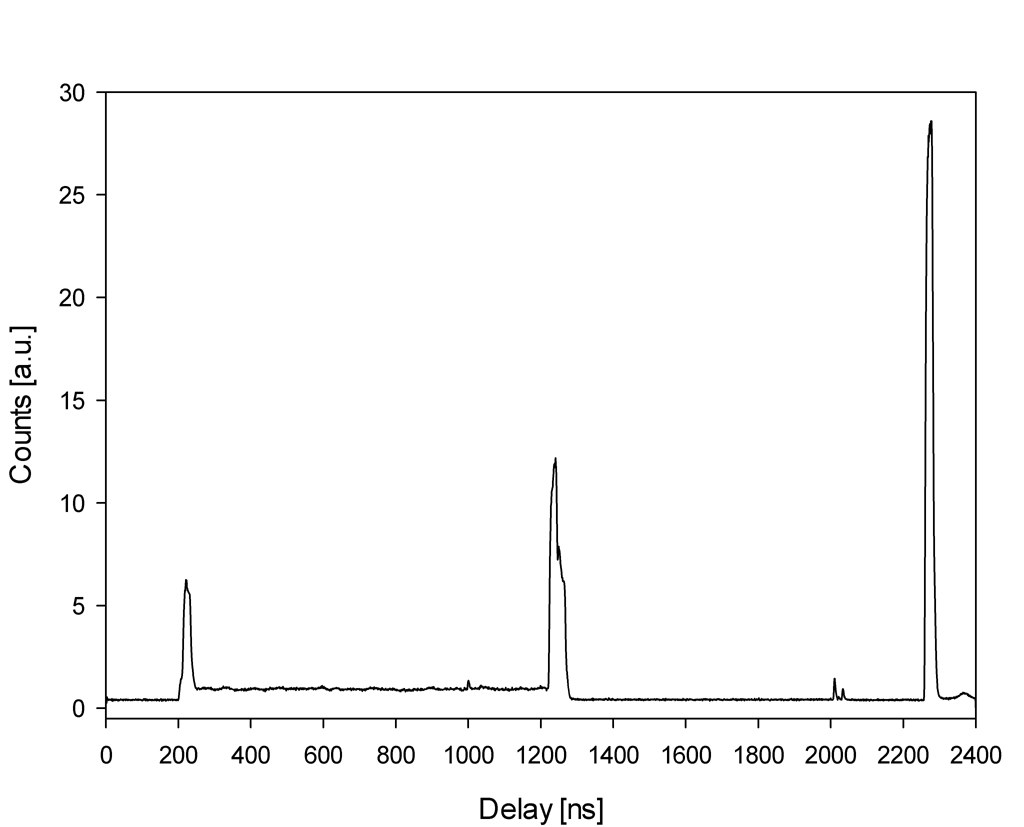}
\caption{Stray photons originating from the down-stream path. The first two peaks are attributed to Alice' and Bob's circulators, respectively, the noise in between is scattering from the fiber and the last peak are the usable photons returning from Bob.}
\label{fig:stray}
\end{center}
\end{figure}

The pulses emerging from the laser-modulator system contain about 0.23 pJ corresponding to 1.8 million photons. Via a circulator they are sent to the fiber network. The finite directivity of the circulator ($\approx 70$ dB) constitutes the first source of stray photons mentioned in section \ref{subsec:packet}. At each fiber connection we get another localized reflection, even though we use angled connectors (APC) exclusively. Additionally there is a distributed backscattering from the whole length of the fiber network due to Rayleigh scattering. Finally, there is a last reflection from the entrance circulator of Bob's unit. All further reflections inside Bob are sufficiently blocked by this circulator. Figure \ref{fig:stray} shows an echo tomography of these contributions recorded with a short packet (4 pulses), 100 meters of fiber simulating the network, and 200 meters total fiber length inside Bob.
\par
From this picture we can already determine the required pulse packet parameters. The region between the second and the third peak is a time of silence, i.e. after the last stray photons have subsided and before the first desired photons arrive from Bob. The pulse packet may have a length $T_p$ so that a minimal gap remains. In other words, the optical length of Bob must be sufficiently large as to contain the whole pulse packet at one time. In practice, one will decide on a certain pulse packet length (also regarding other parameters, such as the detector gate time) and insert delay fibers into Bob as to fulfil the above requirement.
\par
As we demanded, that no stray photons arrive at the detectors together with the useful photons, it becomes clear, that the next pulse packet must not be started before the previous packet has completely returned. This results in a packet repetition period of $T_r = 2 T_f + T_b + T_p$ with the fiber transmission time $T_f$, Bob's storage time $T_b$ and the packet length $T_p$. In practice this will be $T_r = 2 (T_f + T_p)$.
\par
In a real setting the fiber network is laid out as described in section \ref{subsec:arch}. Thus, the above scenario will be a superposition of several network paths, and a compatible time schedule has to be found. This is solved most easily, when the round trip time is on a similar scale for all paths, as mentioned in section \ref{subsec:arch}.

\subsection{Modulation \& Attenuation (Bob)}
\label{subsec:mod}
The whole optical path inside Bob is a loop, which is spanned by the circulator at the entrance. The photons enter to the top in figure \ref{fig:setup} where 99\% are tapped off for generating a clock signal. This clock will be used to drive a pattern generator board, which is currently developed at AIT. In the meantime, we used a direct electronic connection to obtain the necessary clock signal for a standard pulse pattern generator (\textit{70843A Pattern Generator, Hewlett Packard}). This generator cannot produce a random sequence on-the-fly but uses a fixed pseudo random bit sequence (PRBS) repeatedly, for a proof of principle. Next, there is a delay fiber spool in order to provide the necessary storage time of the loop, as mentioned in section \ref{subsec:downstream}. A polarization controller together with a polarizing beam splitter (PBS) ensure the proper operation of the following phase modulator. We use a 10 GHz LiNbO$_3$ phase modulator (\textit{MPZ-LN-10, Photline Technologies}). Before returning to the network, the pulses have to be attenuated to an average photon number around $\mu \approx 0.1$ to ensure security of the protocol (see section \ref{sec:results}). This is done with a variable optical attenuator (VOA) (\textit{MEMS VOA, Agiltron}).

\subsection{Up-stream (PON)}
\label{subsec:upstream}
When returning through the fiber network, the pulses will experience the same loss as in the down-stream path, which is mainly governed by the number of beam splitters depending on the network configuration. In most scenarios this will be the limiting factor for the feasibility of the system, as it determines the finally useful counts at the detectors. When the photons arrive at Alice' entrance, they are routed towards the demodulation unit by the circulator.

\subsection{Demodulation \& Detection (Alice)}
\label{subsec:demod}
The retrieval of the phase information applied by Bob is performed, as described in section \ref{subsec:coding}, by a planar lightwave circuit silicon waveguide Mach-Zehnder interferometer (\textit{M0001NPMSS, NTT Electronics Corporation}). It provides a 1 ns delay (1 GHz free spectral range) and is equipped with a thermo-electric heater. The two output fibers are connected to a single photon detector each (\textit{id200, idQuantique}). By properly adjusting the temperature with respect to the wavelength of the photons, one can achieve a maximum extinction in one exit port for non-modulated pulses. When modulating the pulses alternatingly with zero and $V_\pi$, one can maximize the extinction in the other exit port by optimizing the modulation voltage $V_\pi$.
\par
For each detection event the packet number, the bit number and the detector number is recorded by a multi-channel time tagging unit (\textit{TTM8000, AIT}), which constitutes the raw key data on Alice' side. Since Bob possesses the complete information about all bits in all packets, we have everything ready to be passed on to the post processing software stack.

\section{Results}
\label{sec:results}
\begin{figure}[h!]
\begin{center}
a)
\includegraphics[height=45mm]{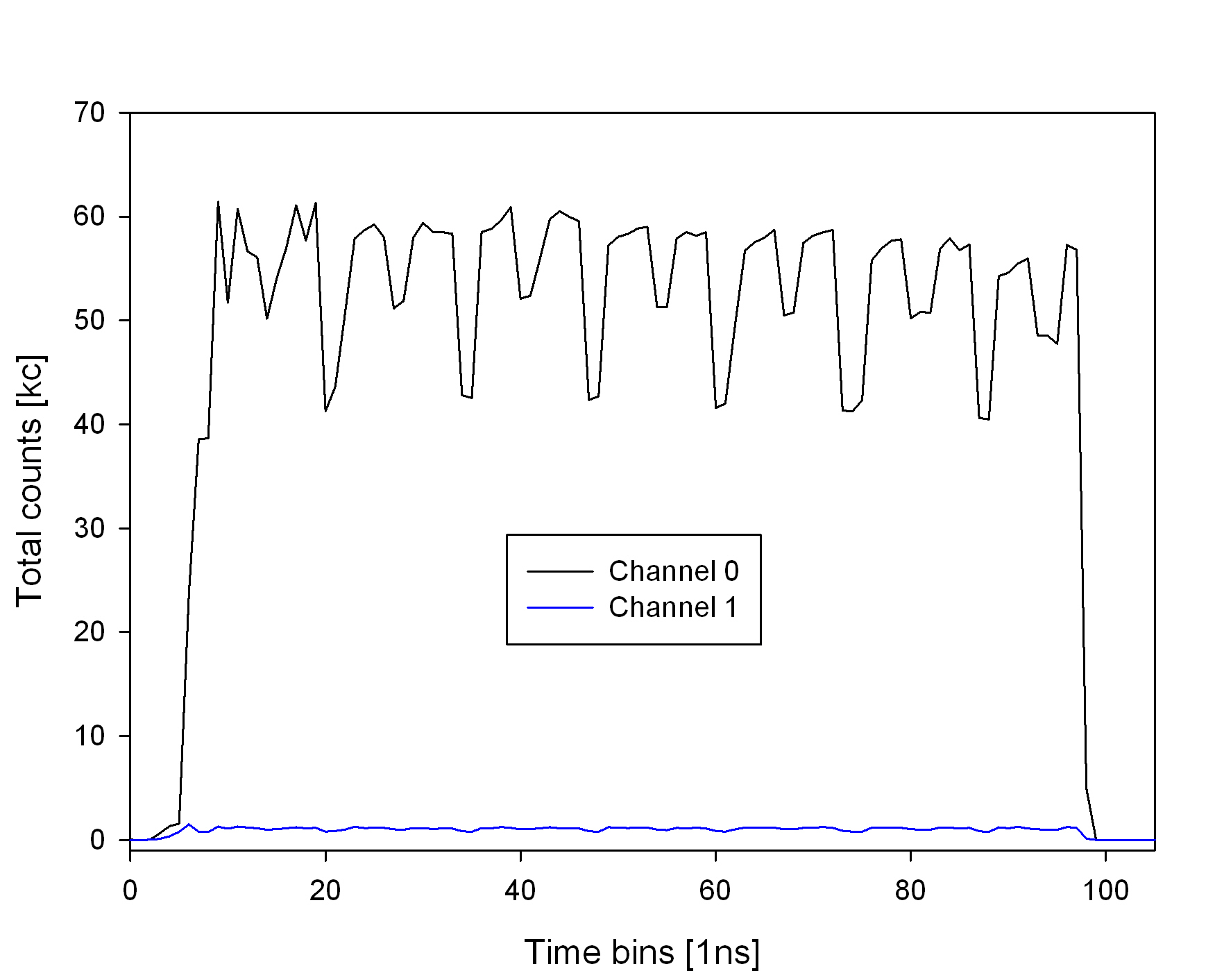}
\includegraphics[height=45mm]{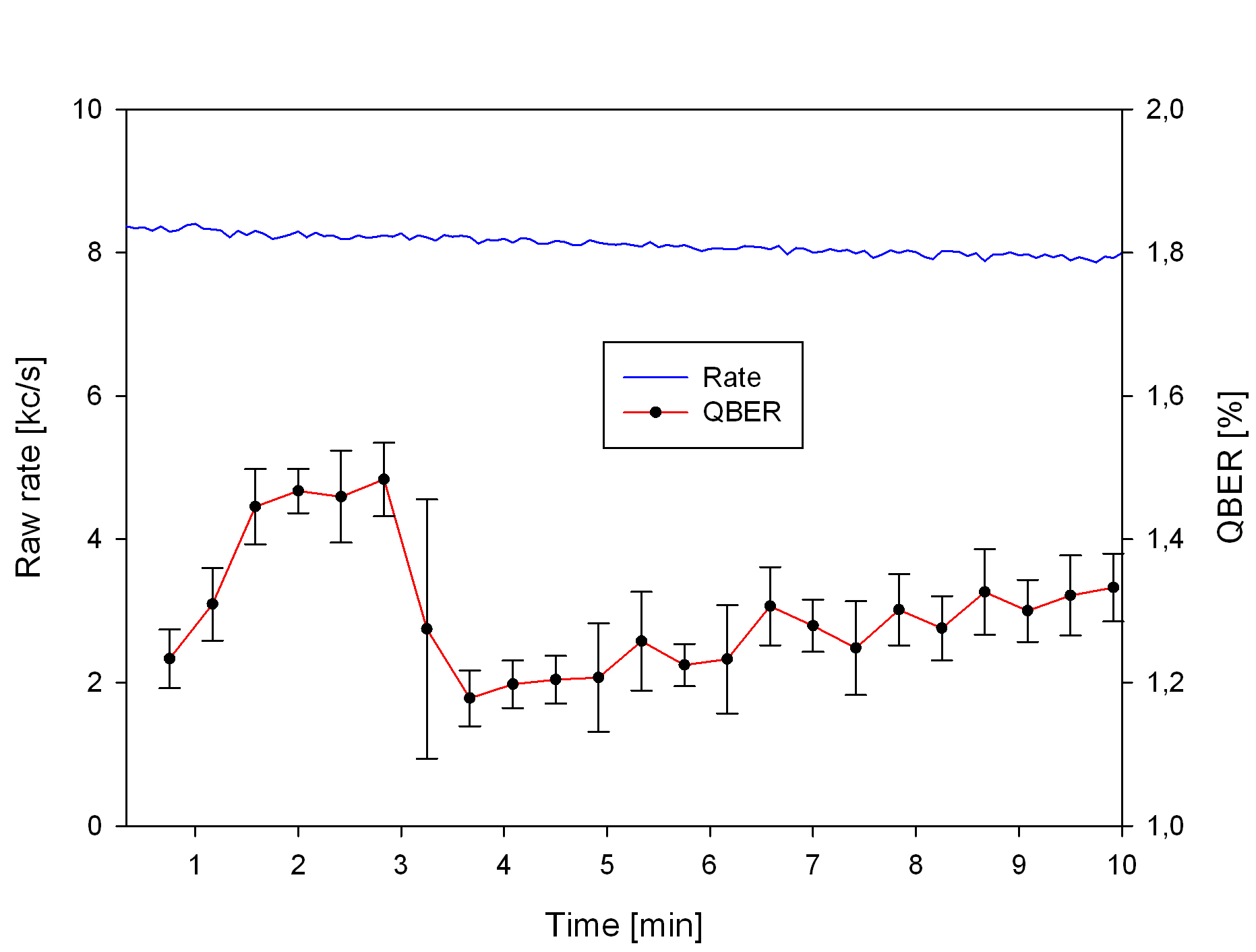}\\
b)
\includegraphics[height=45mm]{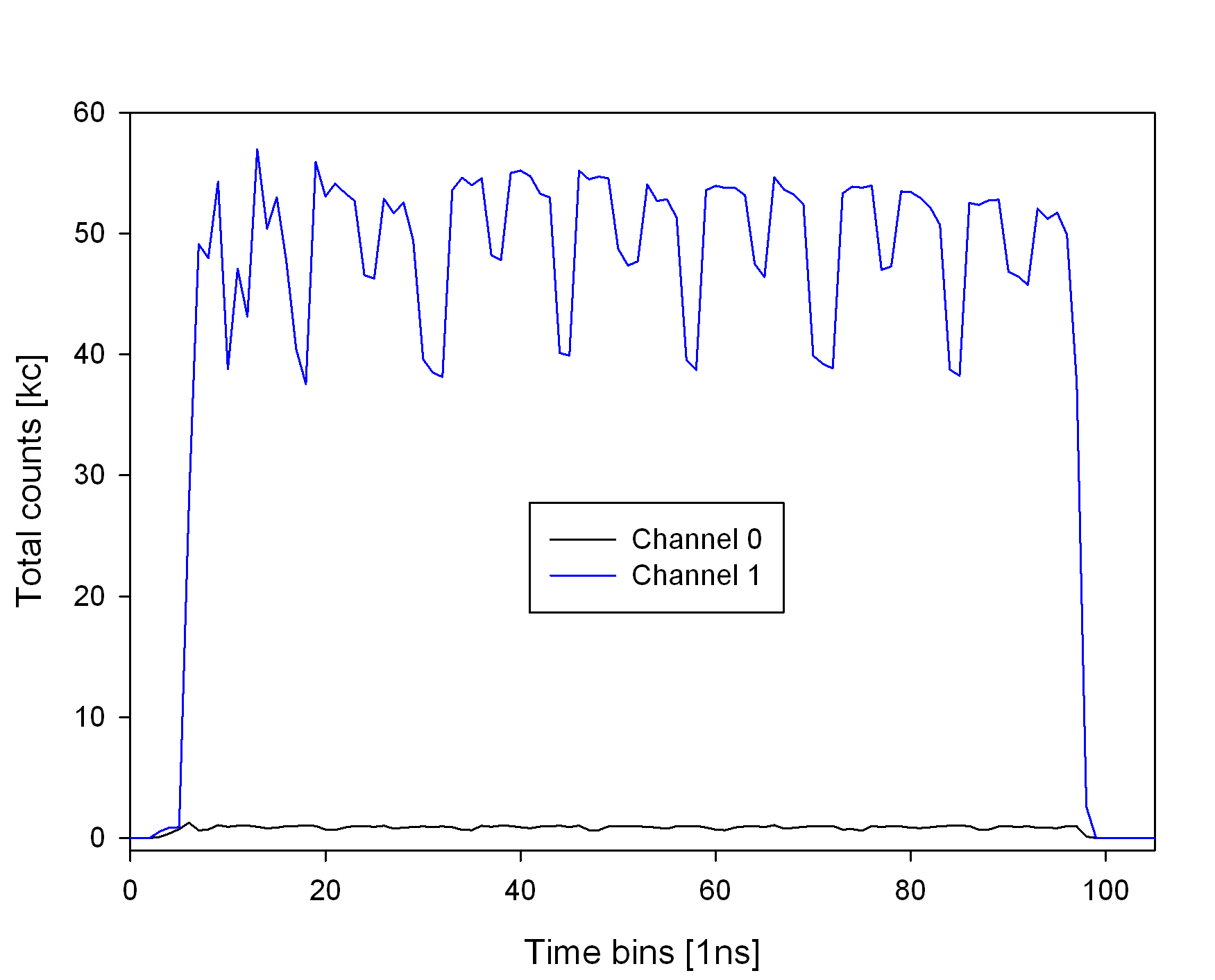}
\includegraphics[height=45mm]{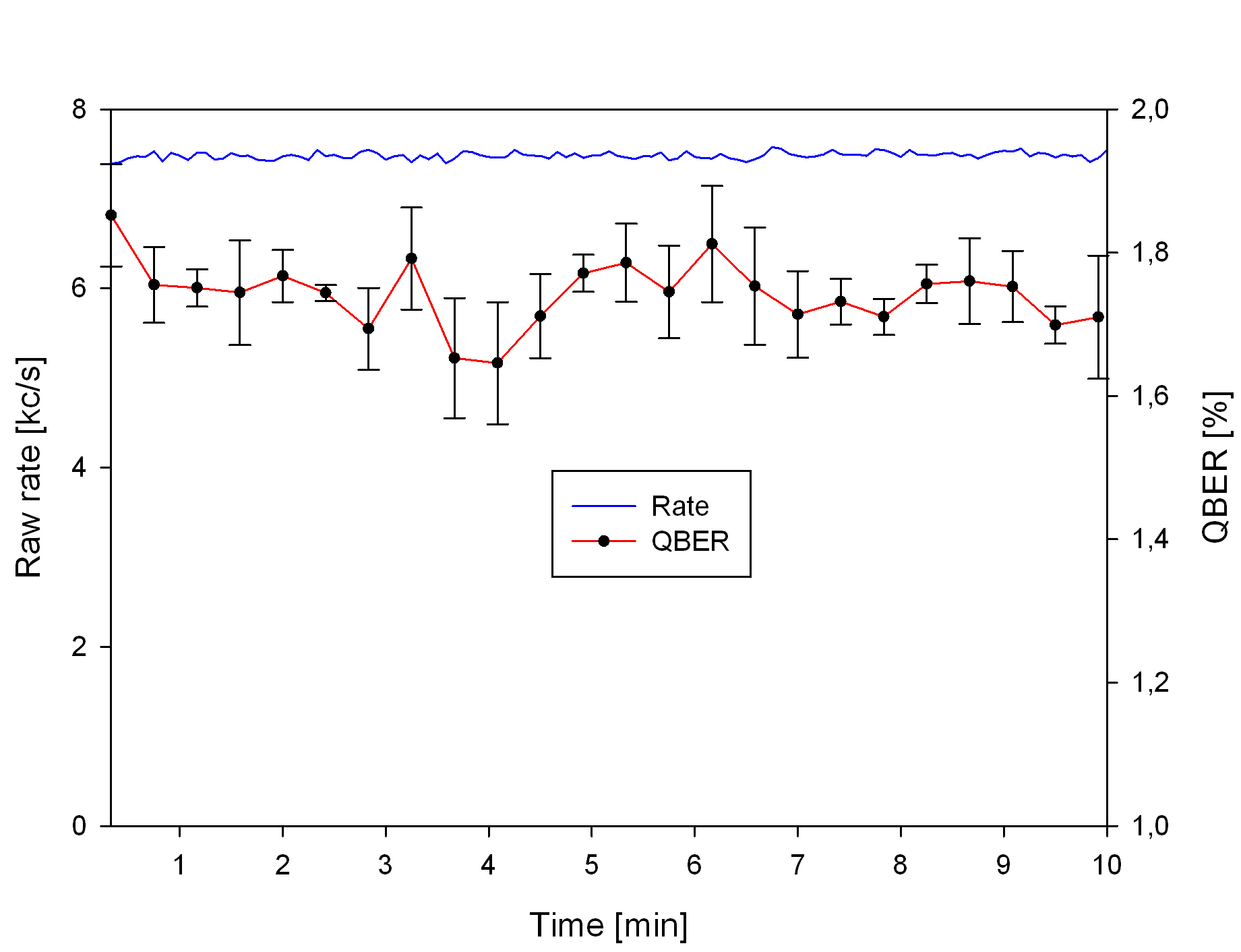}\\
c)
\includegraphics[height=45mm]{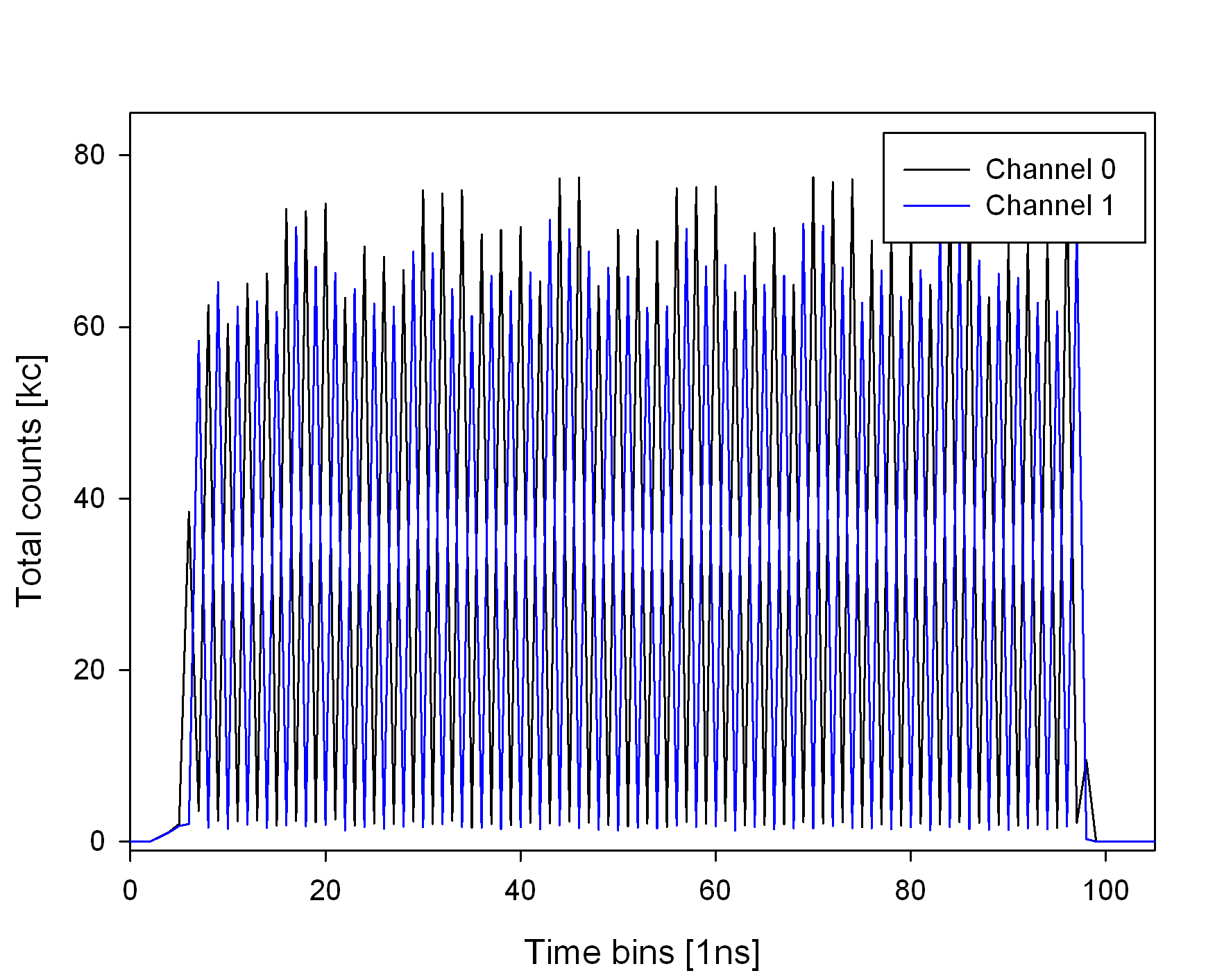}
\includegraphics[height=45mm]{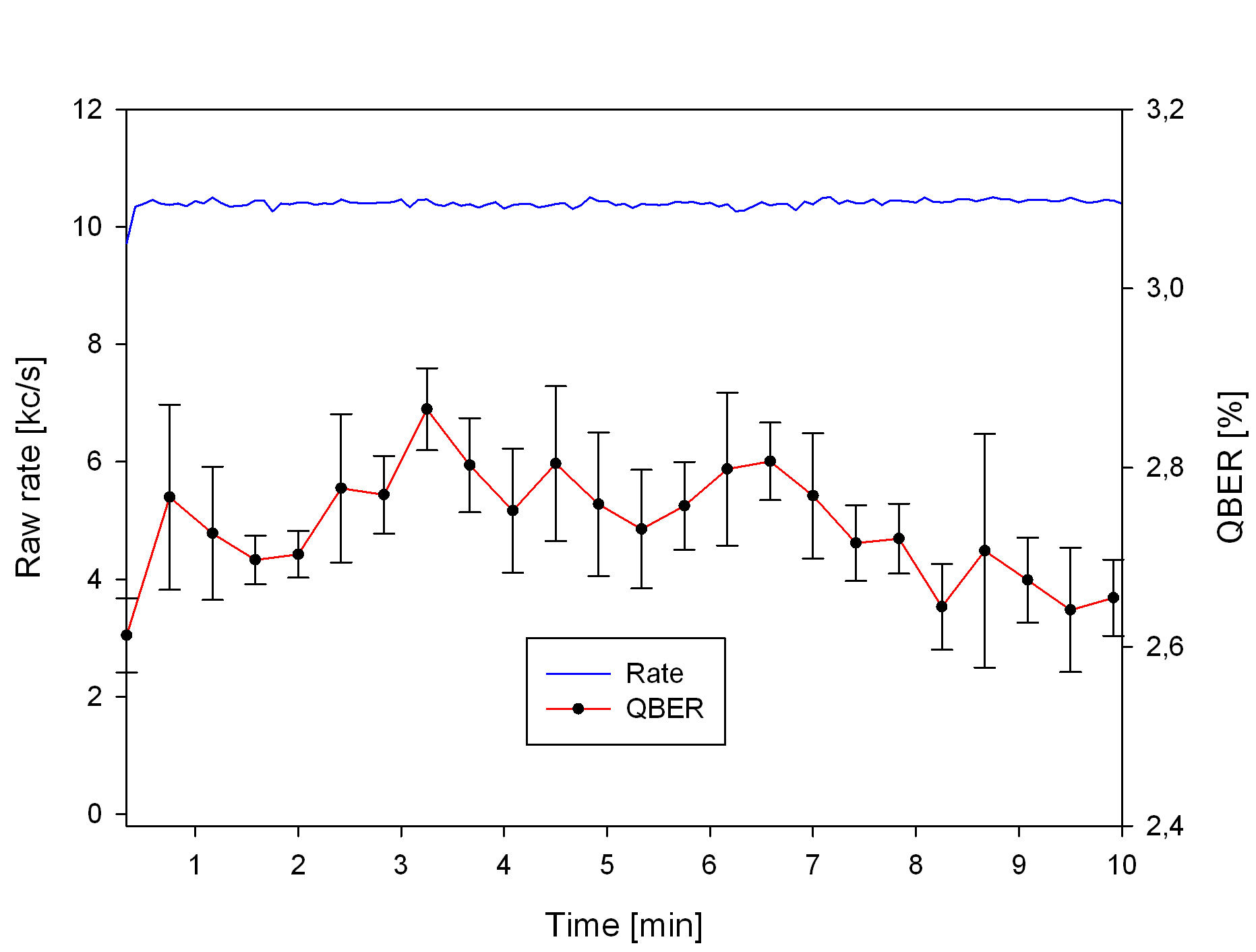}\\
d)
\includegraphics[height=45mm]{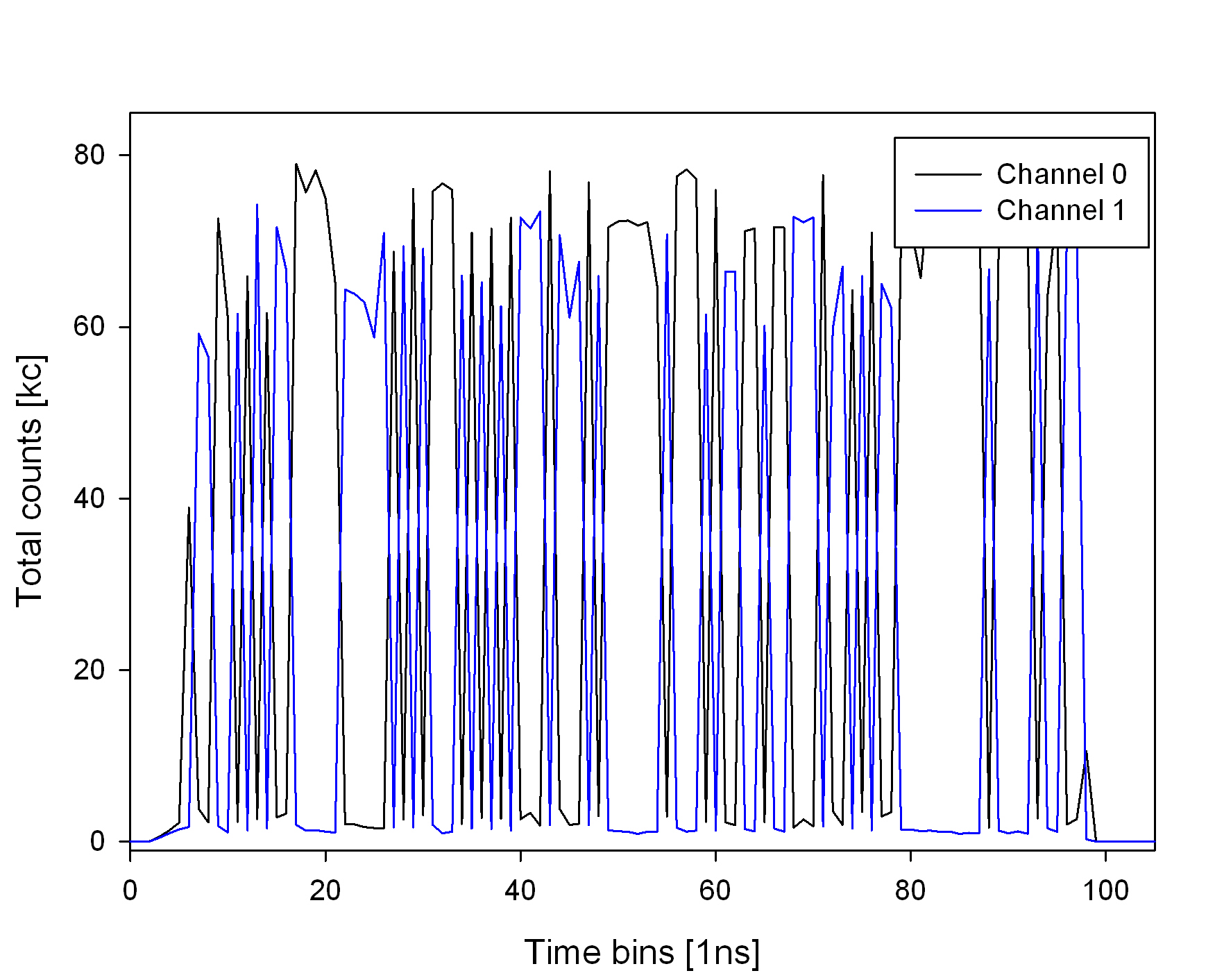}
\includegraphics[height=45mm]{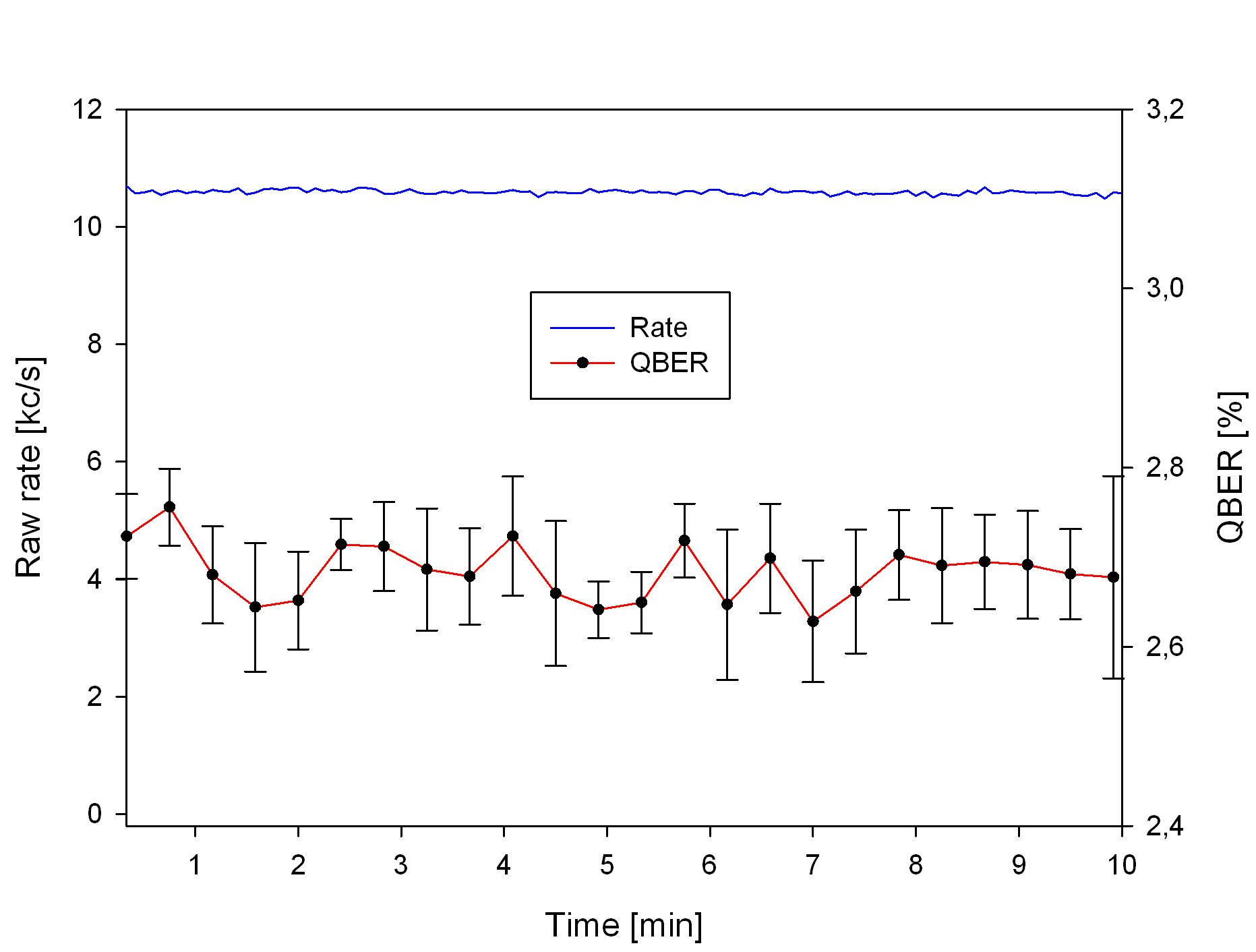}\\
\caption{Test pattern packets accumulated in the detectors (left) and temporal evolution of raw key rate and QBER (right). a) All-zero (no modulation). b) All-one ($\pi$ phase jump for every pulse). c) Alternating ($\pi$ phase jump for every second pulse). d) PRBS (random phase).}
\label{fig:patterns}
\end{center}
\end{figure}

In our demonstration experiment we used the setup shown in figure \ref{fig:setup} with the following parameters: pulse packet length $T_p = 128$ pulses, packet repetition period $T_r = 8192$ ns, network fiber length of 100 m, PON splitting of 1:4 and 200 m storage fiber inside Bob. We should mention, that these parameters are not very well suited, but have rather come about due to various technical reasons (maximum detector gate, available fiber spools, limitations by the electronics) and will be optimized in further experiments. In order to test the basic operation, we connected one Bob unit to the system and applied four different bit patterns instead of a true random key. Figure \ref{fig:patterns} shows the detector histograms, the raw key rates and the QBER for all cases.
\par
Since we have not implemented the complete post-processing stack for secure key distillation, we used theoretical considerations in order to assess the final capabilities of our system. Given a certain raw key rate and QBER we can use existing security proofs in order to estimate the final secure key rate in a realistic scenario. Presently security of DPS QKD is proven only in the case of individual attacks for weak coherent pulses. The corresponding estimates are given in \cite{Waks2006}-\cite{Shimizu2014} (see particularly \cite{Diamanti2006b} for a detailed presentation).
Using the following system parameters average photon number per pulse $\mu = 0.1$, trasmissivity $T = 0.2$, QBER = 2.7\% and overhead of error correction above the Shannon limit of 5\% - using a recently optimized CASCADE algorithm \cite{Martinez2014}, we get the fraction of the raw key rate to remain after post-processing to be approximately 34\% (this ratio is pretty stable with a reduction of T, getting down to 31\% for $T = 10^{-7}$). With a raw key rate of about 10000 counts/s we get 3400 bits/s of final key rate for our system (security against individual attacks as pointed out above).
\par
\begin{figure}[h]
\begin{center}
\includegraphics[height=70mm]{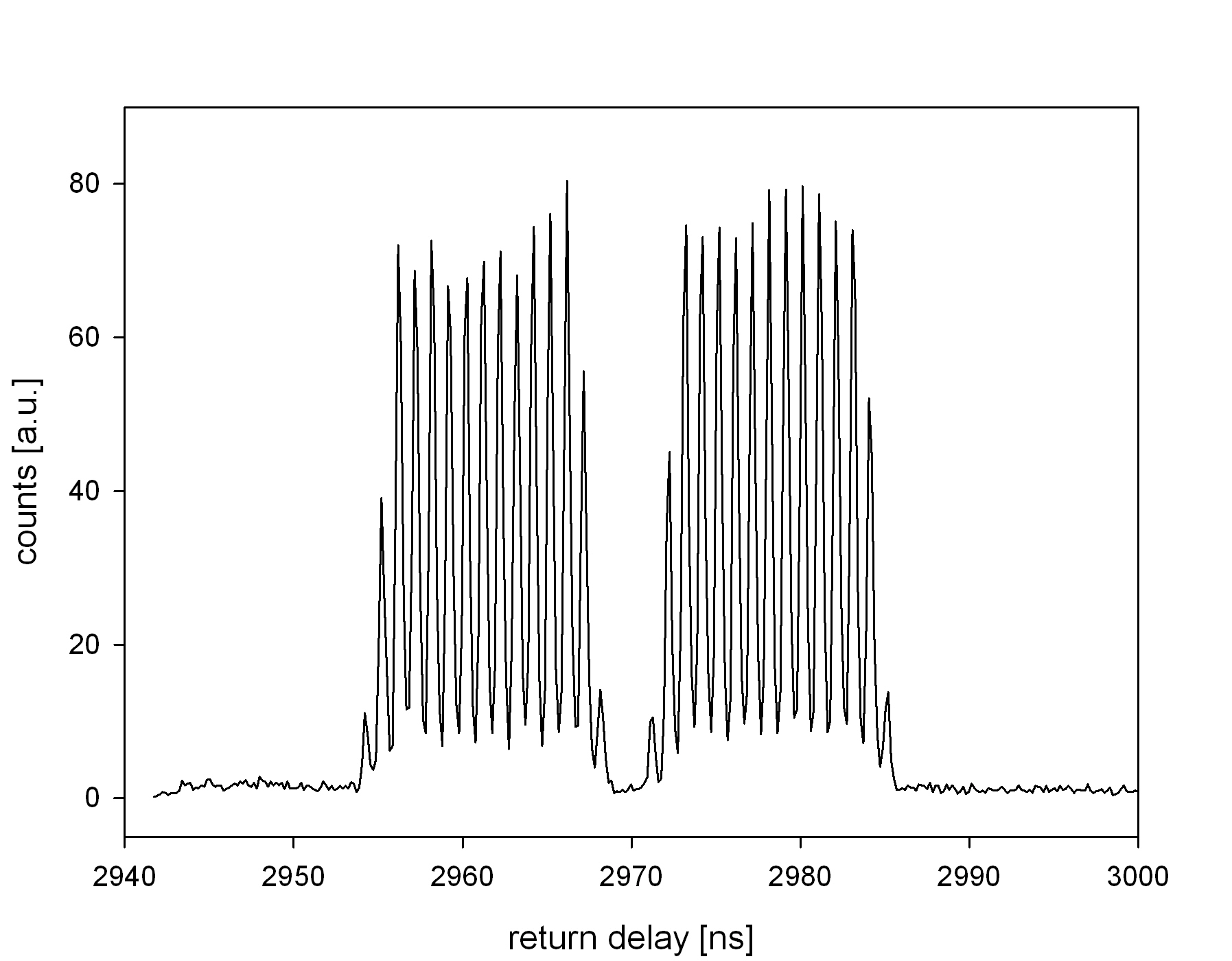}
\caption{Two interleaved packets from different users arriving at Alice.}
\label{fig:multiuser}
\end{center}
\end{figure}

\clearpage
In order to demonstrate a multi-user operation, we connected two Bob units with slightly different storage times to the network and reduced the packet length to 16 pulses. Figure \ref{fig:multiuser} shows the packet histograms at Alice' detectors, which can clearly be distinguished by their different return times with respect to the global departure time.

\section{Conclusions}
\label{sec:conclusions}
We have shown the feasibility of multi-user quantum key distribution in a passive optical network. Our system is based on a differential phase shift scheme and is optimized for cost efficient implementation of network leafs. A flexible configuration of the network is enabled by an interleaved packet scheme, which obviates the need for a particular addressing procedure.

\section*{Acknowledgment}
This work has been supported by the Vienna Science and Technology Fund (WWTF) through project ICT10-048 (LQuNet).

\bibliographystyle{unsrtnat}

\end{document}